# Tachyons as a Consequence of Light-Cone Reflection Symmetry

Alan Chodos[1]

Dept. of Physics, University of Texas at Arlington,
502 Yates Street, Box 19059, Arlington, TX 76019

Abstract:

We introduce a new symmetry, Light-Cone Reflection (LCR), which interchanges timelike and spacelike intervals. Our motivation is to provide a reason, based on symmetry, why tachyons might exist, with emphasis on application to neutrinos. We show that LCR, combined with translations, leads to a much larger symmetry. We construct an LCR-invariant Lagrangian, and discuss some of its properties. In a simple example, we find complete symmetry in the spectrum between tachyons and ordinary particles. We also show that the theory allows for the introduction of a further gauge invariance related to chiral symmetry.

Introduction

The small but energetic community of physicists who study the possibility [1] that neutrinos are tachyons has a variety of issues to deal with. There is the problem of searching existing data, both terrestrial and astrophysical, for hints of tachyonic behavior [2]. There are attempts to explain dark energy and dark matter in terms of tachyonic neutrinos [3]. At a more technical level, there is the question of constructing a sensible quantum theory of neutrinos as tachyons [4].

This paper addresses yet another issue: is it possible to understand the existence of tachyons as the manifestation of some underlying symmetry? After all, gauge bosons must exist in order to implement gauge invariance. And the fermions are arranged in multiplets of the gauge symmetries in the Standard Model. Could neutrinos as tachyons likewise be mandated by the imposition of some sort of symmetry?

We propose a positive answer to this question, in the form of what we call Light-Cone Reflection symmetry (or LCR for short). In this work, we shall define LCR, exhibit a theory of fermions that is LCR-invariant, and discuss some of its properties. But we shall defer discussion of exactly how the theory is to be connected to the Standard Model.

LCR is defined in the context of Very Special Relativity (or VSR for short), as introduced more than 15 years ago by Cohen and Glashow [5]. Specifically we use the sim(2) realization of VSR, which is the maximal possibility. Sim(2) is a proper subgroup of the Lorentz group, which means that the theory we consider is not fully Lorentz invariant. But as Cohen and Glashow explained,

---

[1] alan.chodos@uta.edu

VSR does capture most of the desired ingredients of special relativity, and there have been many papers written since their work that explore the idea that VSR, not the full Lorentz group, is the symmetry actually realized in Nature [6].

We have defined LCR, and studied some of its properties, in eprints that have appeared over the last decade [7-10]. In this paper we incorporate some of that work, and we go on to demonstrate a new result, that LCR, although itself a discrete transformation, imposes a continuous gauge symmetry on the theory, which we explicitly describe.

With the help of auxiliary fields that we call simulons, we can construct a Lagrangian that embodies both VSR and LCR symmetries, incorporating, in addition, the chiral symmetry that was present in Cohen and Glashow's original work on VSR. We find that this Lagrangian has the potential to include a further gauge symmetry, which we show how to implement. We also point out that a particular version of the Lagrangian admits solutions that explicitly exhibit the symmetry between tachyons and ordinary particles that is expected as a result of LCR.

## VSR

In realizing the sim(2) version of VSR, the breaking of the Lorentz group is accomplished by introducing a null 4-vector $n^\mu$, and demanding invariance of the theory under those Lorentz transformations that either leave $n^\mu$ invariant or scale it by a constant factor (the latter are boosts in the $\vec{n}$ direction). It is convenient to choose coordinates so that $n^\mu = (1,0,0,1)$. Our metric is $diag(1,-1,-1,-1)$.

Defining light-cone coordinates $u = n \cdot x = t - z$ and $v = t + z$, we have

$$x^\mu x_\mu = uv - x^2 - y^2. \qquad (1)$$

The 4-parameter sim(2) group of transformations is given explicitly by

(1) Rotations in the x-y plane;
(2) Boosts in the z direction: $u \to \sigma u$;  $v \to \sigma^{-1} v$;
(3) $x \to x + au$;  $v \to v + 2ax + a^2 u$;    and
(4) $y \to y + bu$;  $v \to v + 2by + b^2 u$ .

One sees that $n^\mu$ is unchanged by transformations (1), (3) and (4), and scales by $\sigma^{-1}$ under transformation (2).

Cohen and Glashow used VSR to construct an equation for neutrinos:

$$\left(p_\mu - \frac{m^2}{2n \cdot p} n_\mu\right) \gamma^\mu \psi = 0. \qquad (2)$$

Among its properties is that the usual relativistic dispersion formula, $E^2 - \vec{p}^2 = m^2$, is maintained. In addition, even though the equation describes a massive particle, it possesses a chiral symmetry:

$$\psi \to e^{i\theta\gamma_5}\psi . \qquad (3)$$

The equation is, however, non-local, because of the $n \cdot p$ factor in the denominator. It is not possible to construct a local VSR invariant theory using only fields that transform under representations of the Lorentz group. Below we shall cast the theory in local form by introducing simulon fields, which scale under boosts in the $\vec{n}$ direction.

Although Cohen and Glashow did not mention it, their equation for the neutrino can apply either to ordinary particles ($m^2 > 0$) or to tachyons ($m^2 < 0$). In the latter case, however, there is a caveat: the denominator $2n \cdot p$ can never vanish for $p_\mu$ timelike, but this is no longer true if it is spacelike. Hence, for tachyons, those $p_\mu$ for which $n \cdot p = 0$ must be treated with caution.

LCR

Making use of the vector $n^\mu$, we define the LCR transformation as follows:

$$x^\mu \to x'^\mu = x^\mu - \frac{x^\nu x_\nu}{n \cdot x} n^\mu . \qquad (4)$$

We observe that it has the properties
$$x'^\nu x'_\nu = - x^\nu x_\nu \qquad (5)$$
and

$$\left(x'^\mu\right)' = x^\mu , \qquad (6)$$

inspiring the name Light-Cone Reflection. Under LCR, $n \cdot x = n \cdot x'$. For timelike $x^\mu$, $n \cdot x$ can never vanish, so in fact LCR maps the entire interior of the light cone into its exterior, with the hyperplane $n \cdot x = 0$ removed.

By construction, LCR commutes with the sim(2) transformations. It does not, however, commute with translations. As defined, it is the reflection in a particular light cone, one whose apex is at $x^\mu = 0$. But we want a symmetry that allows reflection in a light cone whose apex is located at an arbitrary spacetime point.

One way to proceed is to compute the commutators of translations with LCR and attempt to make sense of the resulting infinite-dimensional algebra. But this is not necessary, because we can simply infer the result. In order to accommodate translations, LCR must be extended to the following symmetry group:

First, transformations that are connected to the identity, parameterized by a set of functions $\lambda(u, x, y)$:

$$x'^\mu = x^\mu - \lambda(u, x, y)n^\mu \qquad (7)$$

Second, a component parameterized by a set of functions $\kappa(u, x, y)$:

$$x'^\mu = x^\mu - [v + \kappa(u, x, y)]n^\mu. \qquad (8)$$

The original LCR transformation belongs to the second set, with

$$\kappa = -\frac{x^2 + y^2}{u}. \qquad (9)$$

The group multiplication laws are:

1) Two $\lambda$ transformations → a $\lambda$ transformation with parameter $\lambda_1 + \lambda_2$;
2) A $\lambda$ transformation followed by a $\kappa$ transformation → a $\kappa$ transformation with parameter $\kappa - \lambda$;
3) A $\kappa$ transformation followed by a $\lambda$ transformation → a $\kappa$ transformation with parameter $\kappa + \lambda$;
4) Two $\kappa$ transformations → a $\lambda$ transformation with parameter $\kappa_1 - \kappa_2$.

From (4) we note that squaring a $\kappa$ transformation returns the identity, so each $\kappa$ transformation is a reflection of a sort.

A reflection in the light cone with apex at $a^\mu$ can be carried out either by first translating by $a^\mu$ and then making the original LCR transformation, or by making a $\kappa$ type transformation, with parameter

$$\kappa(u, x, y) = \frac{(a^0 + a^3)u + (a^0)^2 - (a^3)^2 - (\vec{a}_\perp + \vec{x}_\perp)^2}{u + n \cdot a}, \qquad (10)$$

where $\vec{a}_\perp = (a^1, a^2)$, $\vec{x}_\perp = (x, y)$, and then translating by $a^\mu$. This indicates that the full set of $\lambda$ and $\kappa$ type transformations is necessary for the symmetry group to close.

Constructing the Lagrangian

We now proceed to construct a Lagrangian that embodies the symmetries we have discussed above. We will not seek to impose the full gauge symmetry, but only the original LCR. However, we expect, and will confirm, that the resulting Lagrangian will be invariant under the full set of gauge transformations.

We begin with a Lagrangian that was put forward by Alvarez and Vidal [11] to yield the Cohen-Glashow VSR-invariant wave equation for the neutrino. As we have noted, this can only be done with the use of auxiliary fields, not all of which transform covariantly under Lorentz transformations.

In our notation, the Alvarez-Vidal Lagrangian reads

$$L = i\bar{\psi}\gamma^\mu \partial_\mu \psi + i\bar{\chi}(n \cdot \partial)\rho + i\bar{\rho}(n \cdot \partial)\chi + \frac{i}{2}m\{\bar{\chi}(n \cdot \gamma)\psi + \bar{\rho}\psi - h.c.\}. \qquad (11)$$

Here $\psi$ represents the neutrino, whereas $\chi$ and $\rho$ are auxiliary fields. $\rho$ is a normal spinor, but $\chi$ is a simulon; it picks up an extra factor of $\sigma$ under boosts in the $\vec{n}$ direction, to compensate for the way that $n^\mu$ scales under those transformations. If one integrates out $\chi$ and $\rho$ one obtains a non-local Lagrangian that yields the Cohen-Glashow equation for $\psi$.

In general, in a term of the Lagrangian in which $n^\mu$ appears, at least one of the fields must be a simulon in order to compensate for the scaling of $n^\mu$ under a boost in the $\vec{n}$ direction. In the case of equation (11), $\psi$ cannot be a simulon because of the first term, so in the mass term we must choose $\chi$ to be the simulon, a choice which also renders the $(\chi\rho)$ terms VSR invariant, assuming that $\rho$ is an ordinary spinor and not a simulon.

$L$ is VSR-invariant, but not LCR invariant, for two reasons: 1) as can easily be seen, $(n \cdot \partial)$ is odd under LCR; and 2) the expression $\gamma^\mu \partial_\mu$ is not invariant under LCR. We shall deal with these in turn.

To compensate for the change in sign in $n \cdot \partial$ we specify that one of $\chi$ and $\rho$ is odd under LCR, and the other even. We choose $\rho$ to be odd. But then the term $\bar{\rho}\psi - \bar{\psi}\rho$ will change sign. We fix this by replacing $\psi$ with a pair of fields, $\psi^{(+)}$ and $\psi^{(-)}$, which are even and odd under LCR respectively, and coupling $\psi^{(+)}$ to $\chi$ and $\psi^{(-)}$ to $\rho$.

To deal with $\gamma^\mu \partial_\mu$, we need to replace $\partial_\mu$ with a "covariant" derivative $D_\mu$. We assume that $\psi^{(\pm)}$ transform simply under LCR: $\psi^{(\pm)}(x) \to \psi^{(\pm)}(x')$ when $x^\mu \to x'^\mu$. Hence the requirement is that $D_\mu(x) \to D_\mu(x')$.

We look for $D_\mu$ in the form

$$D_\mu = \partial_\mu - V_\mu(x) n \cdot \partial \qquad (12)$$

and we see that $V_\mu$ must transform under LCR as

$$V_\mu(x) \to -V_\mu(x') + \partial_\mu \xi(x) \qquad (13)$$

where $\xi(x) = \frac{x^\nu x_\nu}{n \cdot x}$. We want $D_\mu$ to transform as an ordinary vector under VSR, so $V_\mu$ must be a simulon, to compensate for the scaling of $n \cdot \partial$.

This still leaves some latitude in choosing $V_\mu$. The minimal choice is

$$V_\mu = \partial_\mu \phi \qquad (14)$$

with $\phi$ transforming as a scalar simulon under VSR, and as $\phi(x) \to -\phi(x')$ under LCR. It is also subject to the additional constraint (consistent with its simulon nature)

$$n \cdot \partial \phi = 1. \qquad (15)$$

The resulting covariant derivative, $D_\mu^{(0)} = \partial_\mu - \partial_\mu \phi(x)\, n \cdot \partial$, obeys $n^\mu D_\mu^{(0)} = 0$ and $D_\mu^{(0)} \phi = 0$. Because of these properties, there is no possible kinetic energy term for $\phi$ with the requisite symmetries, so $\phi$ is necessarily an auxiliary field that can be eliminated from the equations of motion. Indeed, using $n \cdot \partial = 2\frac{\partial}{\partial v}$, we can implement the constraint $n \cdot \partial \phi = 1$ by writing

$$\phi = \frac{1}{2} v + h(u, x, y) \qquad (16)$$

and then we can change coordinates to

$$v' = v + 2h, \ u' = u, \ x' = x, \ y' = y. \qquad (17)$$

Under this transformation, $D_\mu^{(0)}$ becomes

$$D_0^{(0)} = -D_3^{(0)} = \partial_{u'}; \ D_1^{(0)} = \partial_{x'}; \ D_2^{(0)} = \partial_{y'}, \qquad (18)$$

*i.e.* the covariant derivative becomes an ordinary derivative in the new coordinates.

Another choice for the covariant derivative is to let $V_\mu$ be a more normal vector field, $A_\mu$, with the transformation law given above. Because $(n \cdot \partial)\xi = 2$, we can impose the additional constraint $n^\mu A_\mu = 1$, which will be preserved under LCR. Then the corresponding covariant derivative,

$$D_\mu^{(1)} = \partial_\mu - A_\mu(x) n \cdot \partial \qquad (19)$$

has the property $n^\mu D_\mu^{(1)} = 0$.

To construct a kinetic term for $A_\mu$, one might be tempted to mimic ordinary gauge theory and define

$$\mathcal{F}_{\mu\nu} = D_\mu^{(1)} A_\nu - D_\nu^{(1)} A_\mu. \qquad (20)$$

Under LCR, $\mathcal{F}_{\mu\nu}(x) \to -\mathcal{F}_{\mu\nu}(x')$, so one could imagine using $k\mathcal{F}_{\mu\nu}\mathcal{F}^{\mu\nu}$, where $k$ is some constant, as the kinetic energy term. But under VSR, $\mathcal{F}_{\mu\nu}$ transforms as a simulon, i.e. it scales under boosts in the $\vec{n}$ direction, so the proposed term is not VSR-invariant. In fact, a suitable, if unconventional, kinetic term for $A_\mu$ is

$$k\, (n \cdot \partial) A_\mu\, (n \cdot \partial) A^\mu, \qquad (21)$$

which is LCR invariant because $(n \cdot \partial)\partial_\mu \xi = 0$.

We assemble these ingredients to construct the appropriate generalization of the Alvarez-Vidal Lagrangian. We also generalize the mass term, allowing for both Dirac and Majorana masses. The result is:

$$L = L_\psi + L_{\chi\rho} + L_m + L_{KE} , \qquad (22)$$

where

$$L_\psi = i\bar{\psi}^{(+)}\gamma^\mu D_\mu \psi^{(+)} + i\bar{\psi}^{(-)}\gamma^\mu D_\mu \psi^{(-)} ; \qquad (23)$$

$$L_{\chi\rho} = i\bar{\chi}(n \cdot \partial)\rho + i\overline{\rho}(n \cdot \partial)\chi ; \qquad (24)$$

$$L_m = i\{M_1 \bar{\chi}(n \cdot \gamma)\psi^{(+)} + M_2 \bar{\rho}\psi^{(-)} + M_3^* \bar{\psi}^{(+)}(n \cdot \gamma)\chi^c + M_4^* \bar{\psi}^{(-)}\rho^c - h.c.\} . \qquad (25)$$

Here $D_\mu$ is whichever covariant derivative is chosen to define the theory, either $D_\mu^{(0)}$ or $D_\mu^{(1)}$ or perhaps yet another possibility, and $L_{KE}$ is a kinetic-energy term appropriate to that $D_\mu$. As mentioned above, for $D_\mu^{(0)}$ there is no non-vanishing choice for $L_{KE}$. Thus $\phi$ definitely does not represent any additional degrees of freedom in the theory. For $A_\mu$ in $D_\mu^{(1)}$, we do have the kinetic term of equation (21), but in quantizing a VSR invariant theory, it is natural to use light-cone quantization and to choose $u = n \cdot x$ as the evolution parameter. Since $n \cdot \partial = 2\frac{\partial}{\partial v}$, this so-called kinetic term does not contain derivatives with respect to the evolution parameter, and hence $A_\mu$ is purely an auxiliary field that does not describe new degrees of freedom. One is tempted to conjecture that perhaps $D_\mu^{(0)}$ and $D_\mu^{(1)}$ ultimately describe the same physics, but a proof of this conjecture is beyond the scope of the present work.

As usual, the charge-conjugate field to any fermion field $\psi$ is defined by

$$\psi^c = C\bar{\psi}^T \qquad (26)$$

where $C$ is the charge-conjugation matrix satisfying

$$C^{-1}\gamma^\mu C = -\gamma^{\mu T} . \qquad (27)$$

Properties of the Lagrangian

By construction, $L$ is invariant under both VSR and LCR. It is not difficult to check that $L$ is also invariant under the larger gauge group of $\lambda$ and $\kappa$ transformations. We find that with

$$D_\mu = \partial_\mu - V_\mu(x) n \cdot \partial \qquad (28)$$

the required transformations for $V_\mu$ are

$$V_\mu(x) \to V_\mu(x') - \partial_\mu \lambda(x) \qquad (29)$$

for a $\lambda$ transformation, and

$$V_\mu(x) \to -V_\mu(x') + \partial_\mu[v + \kappa(x)] \qquad (30)$$

for a $\kappa$ transformation. In the case of $D_\mu^{(0)}$, this amounts to $\phi(x) \to \phi(x')$ for a $\lambda$ transformation, and $\phi(x) \to -\phi(x')$ for a $\kappa$ transformation, remembering that $n \cdot \partial \phi = 1$.

$L$ also exhibits a chiral symmetry, in which

$$\psi^{(\pm)} \to e^{i\theta\gamma_5}\psi^{(\pm)} \, ; \, \chi \to e^{i\theta\gamma_5}\chi \, ; and \, \rho \to e^{-i\theta\gamma_5}\rho \, . \qquad (31)$$

If we choose $D_\mu^{(0)}$ as the covariant derivative, then, as remarked above, we can go to special coordinates (essentially replacing $v$ by $2\phi$), denoted by primes, in which $D_\mu^{(0)}$ becomes an ordinary derivative. The equations of motion are then linear, and one can look for exact solutions.

In particular, as described in [8], one can find solutions that depend only on $v'$ and $u'$, which reduces the problem to a 2-dimensional one. Solving the equations of motion, and identifying the d'Alembertian in this 2-dimensional space with the mass-squared operator, one finds the spectrum [8]

$$M^2 = \pm 2\{|A| + |B|\} \, and \, \pm 2\{|A| - |B|\} \, , \qquad (32)$$

where $A$ and $B$ are defined in terms of the mass parameters in the Lagrangian:

$$A = M_2 M_3 - M_1 M_4 \qquad (33)$$

and

$$B = M_3^* M_4 - M_1^* M_2 \, . \qquad (34)$$

The important take-away is that the $M^2$ spectrum exhibits symmetry between positive (non-tachyonic) and negative (tachyonic) values, as might be expected for a theory that possesses LCR invariance. It has not been proved whether this symmetry persists in more complicated realizations of LCR, for example when the derivative $D_\mu^{(1)}$ is used instead of $D_\mu^{(0)}$. But it seems unlikely that tachyons would disappear entirely from the spectrum.

Further Gauge Invariance

As first discussed in [9] and [10], we want to consider transformations of the form

$$\psi_i \to \mathcal{M}(\alpha_i, \beta_i)\psi_i \qquad (35)$$

where

$$\mathcal{M}(\alpha, \beta) = \exp\{\alpha\, n \cdot \gamma + i\beta\, \gamma_5\, n \cdot \gamma\} = 1 + \alpha\, n \cdot \gamma + i\beta\, \gamma_5\, n \cdot \gamma \ . \qquad (36)$$

We have used the fact that $(n \cdot \gamma)^2 = 0$ since $n^\mu$ is a null vector. Here $\alpha$ and $\beta$ are real parameters. In $\psi_i$, the index runs over all four fermi fields in the Lagrangian, i.e. $\psi^{(\pm)}, \chi$ and $\rho$. We begin by allowing each of them to transform with their own parameters $\alpha$ and $\beta$. Invariance of $L_\psi$ under these transformations requires the use of $n \cdot D_\mu = 0$, but imposes no restrictions on the parameters. Invariance of the terms in $L_{\chi\rho}$ requires that $\alpha_\chi + \alpha_\rho = 0$ and $\beta_\chi - \beta_\rho = 0$.

The mass terms containing $\chi$ are automatically invariant, because of the explicit factor of $n \cdot \gamma$. However, we find that the two remaining mass terms, which couple $\psi^{(-)}$ to $\rho$, impose incompatible constraints. The $M_2$ term requires $\alpha_- + \alpha_\rho = 0$ and $\beta_- - \beta_\rho = 0$, whereas invariance of the $M_4^*$ term demands $\alpha_- - \alpha_\rho = 0$ and $\beta_- + \beta_\rho = 0$.

Since these requirements cannot be realized simultaneously, the only way to achieve invariance of the theory under these transformations is to set either $M_2$ or $M_4$ to zero. We have to truncate the mass term given in equation (25).

For definiteness, we set $M_4$ to zero. Having realized invariance under the global transformations, we want now to promote them to gauge transformations.

We do this in the simplest possible manner, by requiring that the gauge parameters be independent of $v$, so that the $L_{\chi\rho}$ term is automatically gauge invariant. The only gauge variance of the Lagrangian then comes from $L_\psi$. To compensate, we introduce a pair of gauge fields, $\tilde{A}_\mu$ and $\tilde{B}_\mu$, and an additional term to the Lagrangian:

$$L_{gauge} = g \sum_{j=\pm} \{i\bar{\psi}^{(j)}\gamma^\mu n \cdot \gamma \psi^{(j)}\tilde{A}_\mu + \bar{\psi}^{(j)}\gamma^\mu \gamma_5 n \cdot \gamma \psi^{(j)}\tilde{B}_\mu\} \ . \qquad (37)$$

Under the transformations described above, but now with the $\alpha$'s and $\beta$'s depending on $(u, x, y)$, we impose the transformations

$$\tilde{A}_\mu \to \tilde{A}_\mu - \frac{1}{g}D_\mu \alpha \ ; \quad \tilde{B}_\mu \to \tilde{B}_\mu + \frac{1}{g}D_\mu \beta \ , \qquad (38)$$

which will render the full $L$ invariant. Since we have introduced only a single pair of gauge fields, we must implement the restrictions $\alpha_+ = \alpha_- = \alpha$ and $\beta_+ = \beta_- = \beta$. We could have included separate gauge coupling parameters for each of the terms, but, for simplicity, we have chosen one overall parameter $g$.

To insure the hermiticity of $L_{gauge}$, we must require

$$n^\mu \tilde{A}_\mu = n^\mu \tilde{B}_\mu = 0, \qquad (39)$$

which are consistent with the gauge transformations since $n \cdot D_\mu = 0$.

$\tilde{A}_\mu$ and $\tilde{B}_\mu$ are simulons, so a suitable kinetic term for them is

$$L_{KE}^{gauge} = k\{(n \cdot \partial)\tilde{A}^\mu (n \cdot \partial)\tilde{A}_\mu + (n \cdot \partial)\tilde{B}^\mu (n \cdot \partial)\tilde{B}_\mu\}, \qquad (40)$$

where $k$ is a conveniently chosen constant.

The generators of these new transformations, $i\, n \cdot \gamma$ and $\gamma_5 n \cdot \gamma$, combine with the generator of the chiral transformation, $\gamma_5$, to form the algebra of rotations and translations in the plane. In order for $L_{gauge}$ to be invariant under a chiral transformation, $\tilde{A}_\mu$ and $\tilde{B}_\mu$ must transform as a chiral doublet.

Conclusions

Our main goal in this paper has been to investigate a symmetry, LCR, that relates tachyons and ordinary subluminal particles. Imposition of this symmetry would then be a way to motivate the appearance of tachyons in Nature.

The most likely application of this idea is to the spectrum of neutrinos. Because LCR is realized in the context of VSR, it would amount to a trade: for neutrinos, we sacrifice full Lorentz invariance, but we gain LCR symmetry plus the gauge symmetry that goes along with it when we include translations. In addition, we see the possibility of further gauge invariance, related to the chiral symmetry, that would qualify as a kind of non-standard neutrino interaction.

In the simple example where the spectrum of the theory was computed, we found that there was symmetry between the tachyons and the ordinary particles, bolstering the expectation that an LCR-invariant theory must contain tachyons. Intriguingly, Ehrlich [12] has found that, at least with our current knowledge of the neutrino masses, a spectrum in which there is symmetry between tachyons and non-tachyons is still experimentally viable. A possible test may come from the KATRIN experiment [13], because the heaviest neutrino in this scenario is of the order of 0.5 eV/c$^2$, if we assume the existence of a sterile neutrino with mass of about 1 eV/c$^2$, for which there is some evidence from short baseline experiments [14].

The major challenge that lies ahead is to combine the dynamics described in this paper with a suitable extension of the Standard Model. It is often said that neutrino masses provide the one definitive case of beyond-the-Standard-Model physics. But neutrino interactions are still described, successfully, in the usual way, via coupling to the $W$ and $Z$ bosons. If LCR is to play a role in the physics of neutrinos, we need to demonstrate how the Lagrangian constructed in this paper can be meshed with the Standard Model into a convincing whole.

Acknowledgments: I am grateful for many conversations with Robert Ehrlich.

References

[1] Chodos, A., A. I. Hauser, and V. A. Kostelecký: The neutrino as a tachyon. *Phys. Lett. B* **1985**, *150B*, 6, 431-435.

[2] Ehrlich, R.: A Review of Searches for Evidence of Tachyons, *Symmetry* 2022, 14(6), 1198; Jentschura, U. and R. Ehrlich: Lepton Pair Čerenkov Radiation Emitted by Tachyonic Neutrinos: Lorentz-Covariant Approach and IceCube Data. *Adv. High Energy Phys.* **2016**, *2016*, 4764981; Ehrlich, R. Six observations consistent with the electron neutrino being a tachyon with mass: $m^2 = -0.11 \pm 0.016$ eV$^2$ or $|m| = 0.33 \pm 0.024$ eV. *Astropart. Phys.* **2015**, *66*, 11.

[3] Schwartz, Charles: A Consistent Theory of Tachyons with Interesting Physics for Neutrinos, *Symmetry* **2022**, *14*(6) 1172.

[4] Rembieliński, Jakub, Paweł Caban and Jacek Ciborowski: Quantum Field Theory of Space-like Neutrino, *Eur. Phys. J.* C 81 (2021) 716 ; Radzikowski, M.: A Quantum Field Model for Tachyonic Neutrinos with Lorentz Symmetry Breaking, arXiv:1007.5418 ; Jentschura, U. D, and B. J. Wundt: Localizability of Tachyonic Particles and Neutrinoless Double Beta Decay, *Eur. Phys. J.* C 72 (2012) 1894 ; Schwartz, Charles, reference [3].

[5] Cohen, A. G. and S. L. Glashow: Very Special Relativity, *Physical Review Letters* 97, 021601 (2006). ; Cohen, A. G. and S. L. Glashow: A Lorentz-violating Origin of Neutrino Mass?, arXiv preprint hep-ph/0605036 (2006).

[6] See, for example: Cohen, A. G. and D. Z. Freedman: SIM(2) and SUSY, *Journal of High Energy Physics* 2007, 039 (2007) ; Dunn, A. and T. Mehen: Implications of SU (2)L × U (1) Symmetry for SIM(2) Invariant Neutrino Masses, arXiv preprint hep-ph/0610202 ; Kouretsis, A., M. Stathakopoulos, and P. C. Stavrinos: General Very Special Relativity in Finsler Cosmology, *Phys. Rev. D* 79, 104011 (2009) ; Alfaro, Jorge; Pablo González, and Ricardo Ávila: Electroweak Standard Model with Very Special Relativity, *Phys. Rev. D* 91, 105007 (2015) ; Koch, Benjamin; Enrique Muñoz, and Alessandro Santoni: Corrections to the Gyromagnetic Factor in Very Special Relativity, arXiv:2208.09824.

[7] Chodos, A.: Light Cone Reflection and the Spectrum of Neutrinos, arXiv:1206.5974 .

[8] Chodos, A.: A Model of Neutrinos, arXiv:1511.06745 .

[9] Chodos, A.: Gauge Interactions and a Quantum Avatar in a Model with Light Cone Reflection Symmetry, arXiv:1603.07053 .

[10] Chodos, A.: Chirally Invariant Avatar in a Model of Neutrinos with Light Cone Reflection Symmetry, arXiv:1609.01378 .

[11] Alvarez, E. and R. Vidal : Very Special (de Sitter) Relativity, *Phys. Rev. D* **77**, 127702 (2008).

[12] Ehrlich, R.: A Review of Searches for Evidence of Tachyons, *Symmetry* 2022, 14(6), 1198, section 10.

[13] Aker, M. et al. (KATRIN Collaboration): Direct Neutrino-Mass Measurement with Sub-electronvolt Sensitivity, *Nature Physics* 18 (2) 2022.

[14] For a review of the situation concerning the short-baseline anomalies, see Abdallah, Waleed; Raj Gandhi; and Samiran Roy: Requirements on common solutions to the LSND and MiniBooNE excesses: a post-MicroBooNE study, *JHEP* 06, 160 (2022).